\documentclass[%
 reprint,onecolumn,
 amsmath,amssymb,
 aps,nofootinbib,
]{revtex4-2}
\usepackage{fancyhdr}
\usepackage[colorlinks]{hyperref}
\usepackage[nameinlink,noabbrev]{cleveref}
\hypersetup{
    colorlinks=true,
    linkcolor=blue,
    filecolor=magenta,      
    urlcolor=blue,
   citecolor=blue
}
\usepackage{graphicx}
\usepackage{dcolumn}
\usepackage{bm}

\usepackage{scalerel}
\usepackage{tikz}
\usetikzlibrary{svg.path}

\definecolor{orcidlogocol}{HTML}{A6CE39}
\tikzset{
  orcidlogo/.pic={
    \fill[orcidlogocol] svg{M256,128c0,70.7-57.3,128-128,128C57.3,256,0,198.7,0,128C0,57.3,57.3,0,128,0C198.7,0,256,57.3,256,128z};
    \fill[white] svg{M86.3,186.2H70.9V79.1h15.4v48.4V186.2z}
                 svg{M108.9,79.1h41.6c39.6,0,57,28.3,57,53.6c0,27.5-21.5,53.6-56.8,53.6h-41.8V79.1z M124.3,172.4h24.5c34.9,0,42.9-26.5,42.9-39.7c0-21.5-13.7-39.7-43.7-39.7h-23.7V172.4z}
                 svg{M88.7,56.8c0,5.5-4.5,10.1-10.1,10.1c-5.6,0-10.1-4.6-10.1-10.1c0-5.6,4.5-10.1,10.1-10.1C84.2,46.7,88.7,51.3,88.7,56.8z};
  }
}

\newcommand\orcidicon[1]{\href{https://orcid.org/#1}{\mbox{\scalerel*{
\begin{tikzpicture}[yscale=-1,transform shape]
\pic{orcidlogo};
\end{tikzpicture}
}{|}}}}
\usepackage{amsmath}
\usepackage{amssymb, amsfonts}
\usepackage{babel}
\begin{document}

\preprint{APS/123-QED}

\title{Gravitational instability with a dark matter background: Exploring the different scenarios}

\author{Kamel Ourabah \orcidicon{0000-0003-0515-6728},}\email{kam.ourabah@gmail.com}
\address{Theoretical Physics Laboratory, Faculty of Physics, University of Bab-Ezzouar, USTHB, Boite Postale 32, El Alia, Algiers 16111, Algeria}




\date{\today}

\begin{abstract}
We study the Jeans-type gravitational instability for a self-gravitating medium composed of two species, baryonic (bright) and dark matter particles, using a hybrid quantum-classical fluid approach. Baryonic matter is treated classically, which is appropriate for most astrophysical environments, e.g., Bok globules, while dark matter is treated through a quantum hydrodynamic approach allowing for possible nonlinearities. These nonlinearities may arise in bosonic dark matter due to attractive or repulsive short-range self-interaction (attractive interaction being more relevant for axions) or from the Pauli exclusion principle for fermionic dark matter, e.g., massive neutrinos. This allows us to explore, in a very broad context, the impact of a dark matter background on the Jeans process for different scenarios discussed in the literature. 
We confront the established stability criterion with Bok globule stability observations and show that the model adequately accounts for the data with dark matter parameters close to those predicted independently from numerical simulations. 
\end{abstract}

\maketitle

\section{Introduction}
Explaining the transition from the nearly homogeneous and isotropic early Universe, as revealed by the cosmic microwave background radiation, to the highly clustered Universe observed today, constitutes one of the major issues in cosmology.
The first theory explaining the formation of structures through gravity dates back as early as 1902, in a paper by Jeans \cite{Jeans}. Assuming that the Universe is filled with a fluid, Jeans successfully combined hydrodynamic equations with Newtonian gravity to address the formation of self-gravitating objects by means of the interplay between the gravitational attraction and the pressure forces acting against it. We are 120 years later now and the Jeans instability still stands as an actively studied physical problem, revisited from various modern standpoints. One may cite for example its extension to general relativity \cite{JGR} and to an expanding universe background \cite{JGR,B,G}, its formulation in the language of kinetic theory \cite{Jkin1,Jkin2,Jkin3}, its generalization to alternative gravity theories \cite{Jmod1,Ahmed,Jmod2,Jmod3}, where it regularly serves as a possible phenomenon to constraint their parameters \cite{Vainio,Claudio}. The Jeans mechanism has also been extensively studied in various dark matter models \cite{Boyanovsky,Chavanis2011,Ourabah2020,Ourabah2020bis,Chavanis2020} and for mixtures of baryonic and dark matter particles \cite{Jkin3,Kremer1}. 

When it comes to dark matter, one faces a serious limitation; little is known so far about
the nature of dark matter, which leads to a wide range of theoretical speculations with dramatically different consequences. In the $\Lambda$CDM standard cosmological model, dark matter is thought to be cold and is depicted as a classical pressureless gas. This
model works extremely well at large (read cosmological) scales, and can explain measurements
of the cosmic microwave background radiation \cite{Ade}. At small (read galactic) scales however, it is a different story. The model suffers from a number of serious drawbacks; the most well-documented being known as the core-cusp problem \cite{crisis1}, the missing satellites problem \cite{crisis2}, and the too-big-to-fail problem \cite{crisis3} (see Ref. \cite{crisis4} for a review on the small-scale $\Lambda$CDM crisis).

Various proposals have been put forward to fix these small-scale problematic aspects, either by staying within the cold dark matter (CDM) paradigm but invoking the feedback of baryons \cite{s1} or collisions between particles \cite{s2}, or by discarding the assumption of a pressureless fluid, considering warm dark matter particles \cite{crisis1,Warm}. In this case, the interplay between the gravitational
attraction and the velocity dispersion (or temperature) is expected to resolve the aforementioned issues. Another possibility is to take into
account quantum mechanics, which is known to produce an effective pressure even at zero temperature. Quantum effects being relevant only at small length scales, they may solve the small scale CDM crisis (for dark matter particles with a sufficiently small mass), while they fade away at larger scales, maintaining the virtues of the CDM model at the cosmological scale. 

Here again, various scenarios have been suggested in the literature. One possibility is to consider that the dark matter particle is a fermion, e.g., a massive
neutrino (see e.g., \cite{Ch19} and references therein). In that case, the gravitational interaction is balanced by the quantum pressure, ultimately produced by the Heisenberg uncertainty principle, and an additional pressure term accounting for the Pauli
exclusion principle (like in white dwarf stars). Another possibility is to consider that the dark matter particle is a boson, e.g., an ultralight
axion (see e.g., \cite{t1,t2}). At zero temperature, bosons form a Bose–Einstein condensate (BEC); This leads to the BECDM model \cite{t1}. In that case, the situation is similar to what happens in boson stars; the gravitational attraction is balanced by the quantum pressure, accounting for the Heisenberg uncertainty principle and, possibly, additional pressure forces arising from scattering, if the bosons are self-interacting. 

Here, we wish to address, in a very broad context, the problem of Jeans gravitational instability for a mixture of baryonic and dark matter particles, at the fluid level of description. Wile baryonic matter is treated classically, which is well-motivated for most astrophysical situations, dark matter is treated through a quantum fluid model, constructed \textit{such that} it can accommodate various dark matter candidates. Trough this model, we establish the Jeans criterion for stability and discuss its limits in various scenarios. We compare our predictions to the observed stability of Bok globules and show that the model adequately accounts for the data for dark matter parameters close to those predicted by independent numerical simulations. 

As we are mainly interested in the Jeans criterion, we
shall consider the case of a static universe. As known, the
Jeans wave number can be correctly extracted in the non-expanding
case; one may still include the redshift dependence \textit{a posteriori} \cite{Boyanovsky}. Besides, although we restrict ourselves in the main text to the fluid level of description, the problem can be equivalently addressed in the language of kinetic theory, by employing a mix of Boltzmann and Wigner equations. This will be shown in the Appendix.

\section{The model}

To begin with, let us explicitly lay out the model. We aim at a description of the Jeans mechanism for a mixture of baryonic and dark matter particles, in the weak-field (Newtonian) regime, and at the fluid level of description. Baryonic matter will be treated classically through a hydrodynamic approach. That is,

\begin{equation}\label{1}
\begin{aligned}
\frac{\partial \rho}{\partial t}+\nabla \cdot(\rho \mathbf{u}) &=0,\\
\frac{\partial \mathbf{u}}{\partial t}+(\mathbf{u} \cdot \nabla) \mathbf{u}&=-\frac{\nabla p}{\rho}-\nabla \Phi,
\end{aligned}
\end{equation}
where $\rho$ and $\mathbf{u}$ denote, respectively, the mass density and the fluid velocity field, while $\Phi$ is the gravitational potential, satisfying the Poisson equation. The first equation in (\ref{1}) is the continuity equation and the second one is the Euler (momentum balance) equation. Above, $p$ stands for the pressure; it is linked to the density $\rho$ through an equation of state. We consider here the general case of a polytropic equation of state
\begin{equation}\label{poly}
p=K \rho^{\gamma}, \quad \gamma=1+\frac{1}{n},
\end{equation}
where $K$ is a constant and $n$ is known as the \textit{polytropic index}.

For dark matter particles, we consider a more general fluid model to account for various particle candidates. We will start by modeling dark matter as a self-gravitating BEC (i.e., BECDM) potentially involving short-range interactions but, as will become clear next, such a model covers equally well fermionic and classical dark matter particles in the proper limit. At zero temperature, all the bosons are expected to form a BEC and the system is described by the condensate wave function $\psi(\mathbf{r},t)$. In the \textit{mean-field approximation}, the ground state properties of the BEC are described by the Schrödinger equation

\begin{equation}\label{GP}
i \hbar \frac{\partial \psi}{\partial t}(\mathbf{r}, t)=-\frac{\hbar^{2}}{2 m} \Delta \psi(\mathbf{r}, t)+m \Phi_{\operatorname{tot}}(\mathbf{r}, t) \psi(\mathbf{r}, t), 
\end{equation}
with
\begin{equation}\label{cond}
\begin{gathered}
\int|\psi(\mathbf{r}, t)|^{2} d \mathbf{r}=1, \\
\rho(\mathbf{r}, t)=N m|\psi(\mathbf{r}, t)|^{2} , \\
\Phi_{\operatorname{tot}}(\mathbf{r}, t)=\int \rho\left(\mathbf{r}^{\prime}, t\right) u\left(\left|\mathbf{r}-\mathbf{r}^{\prime}\right|\right) d \mathbf{r}^{\prime}.
\end{gathered}
\end{equation}
Eq. (\ref{GP}) governs the evolution of the wave function $\psi(\mathbf{r},t)$ wile the conditions (\ref{cond}) account for the normalization of the wave function, the density of the condensate, and the associated potential $\Phi_{tot} \equiv \Phi + \Phi_{coll}$. In the general case, this potential refers both to the long-range gravitational potential $\Phi$ and to (possible) short-range interactions, i.e., binary collisions, that can be modeled through a pair contact potential
$u_{S R}\left(\mathbf{r}-\mathbf{r}^{\prime}\right)=g \delta\left(\mathbf{r}-\mathbf{r}^{\prime}\right)$ where $\delta$ is the Dirac delta distribution and the coupling constant $g$ reads as $g=4 \pi a \hbar^{2} / m^{3}$, {$\hbar$ being the reduced Planck constant and} $a$ is the s-wave scattering length \cite{gpp}. Here, two cases have to be distinguished; $a>0$ ($g>0)$ corresponds to repulsive short-range interactions whereas $a < 0$ ($g<0$) corresponds to attractive interactions. In these conditions, the effective potential modeling the short-range interaction (collisions) reads as

\begin{equation}\label{h}
\Phi_{coll}(\rho)=g N m|\psi|^{2}=g \rho
\end{equation}
and, accounting for this potential, the Schrödinger equation reads

\begin{equation}\label{GP2}
i \hbar \frac{\partial \psi}{\partial t}=-\frac{\hbar^{2}}{2 m} \Delta \psi+m(\Phi+ \Phi_{coll}(\rho)) \psi,
\end{equation}
i.e., the Gross-Pitaevskii equation \cite{gpp}. As known, the latter can be written in the form of a hydrodynamic set of equations using the so-called Madelung (or better Madelung-de Broglie-Bohm) transformation \cite{Madelung}. One start by writing the wave function in polar form

\begin{equation}\label{polar}
\psi(\mathbf{r}, t)=  A(\mathbf{r}, t) e^{i S(\mathbf{r}, t) / \hbar},
\end{equation}
where $A(\mathbf{r}, t)$ and $S(\mathbf{r}, t)=(\hbar / 2 i) \ln \left(\psi / \psi^{*}\right)$ are real functions, representing the amplitude and the phase of the wave function. The mass density and the velocity field are defined in terms of $A$ and $S$ as \cite{Madelung}

\begin{equation}
\rho(\mathbf{r}, t)= N m |\psi|^{2}= N m A(\mathbf{r}, t)^2 \quad \text { and } \quad \mathbf{u}=\frac{\nabla S}{m }=\frac{i \hbar}{2 m } \frac{\psi \nabla \psi^{*}-\psi^{*} \nabla \psi}{|\psi|^{2}}.
\end{equation}
Note that, so defined, the velocity field is irrotational, i.e., $\nabla \times \mathbf{u}=\mathbf{0}$. Substituting the wave function (\ref{polar}) into the Gross-Pitaevskii equation (\ref{GP2}) and splitting apart the real and imaginary parts, one has
\begin{equation}\label{8}
\begin{gathered}
\frac{\partial \rho}{\partial t}+\nabla \cdot(\rho \mathbf{u})=0, \\
\frac{\partial S}{\partial t}+\frac{1}{2 m}(\nabla S)^{2}+m \Phi+m \Phi_{coll}(\rho)+Q=0,
\end{gathered}
\end{equation}
where
\begin{equation}
Q \equiv -\frac{\hbar^{2}}{2 m} \frac{\Delta \sqrt{\rho}}{\sqrt{\rho}}=-\frac{\hbar^{2}}{4 m}\left[\frac{\Delta \rho}{\rho}-\frac{1}{2} \frac{(\nabla \rho)^{2}}{\rho^{2}}\right]
\end{equation}
is known as the quantum potential. The first equation in (\ref{8}) is the continuity equation while the second one is the quantum extension to the Hamilton-Jacobi equation, affected by the so-called quantum potential $Q$. Taking the gradient of the Hamilton-Jacobi equation (and noting that $\nabla \times \mathbf{u}=\mathbf{0}$), one ends up with an Euler-like equation
\begin{equation}
\frac{\partial \mathbf{u}}{\partial t}+(\mathbf{u} \cdot \nabla) \mathbf{u}=-\nabla \Phi-\nabla \Phi_{coll}-\frac{1}{m} \nabla Q,
\end{equation}
or, equivalently, as
\begin{equation}
\frac{\partial \mathbf{u}}{\partial t}+(\mathbf{u} \cdot \nabla) \mathbf{u}=-\nabla \Phi -\frac{1}{\rho} \nabla p-\frac{1}{m} \nabla Q
\end{equation}
where the pressure $p(\mathbf{r},t) \equiv p(\rho(\mathbf{r},t))$ is a a function of the density $\rho(\mathbf{r},t)$ (since $\Phi_{coll}$ is a function of $\rho$). For a potential $\Phi_{coll}$ in the form of Eq. (\ref{h}), the pressure reads as
\begin{equation}
p=\frac{2 \pi a \hbar^{2}}{m^{3}} \rho^{2}.
\end{equation}
This corresponds to a polytropic equation of state (\ref{poly}) with 
\begin{equation}
K= \frac{2 \pi a \hbar^{2}}{m^{3}}  \quad \text{and} \quad \gamma=2 \quad (n=1).
\end{equation}
That is, the GP equation is equivalent to a quantum hydrodynamic model with a polytropic equation of state with index $n=1$. 

Interestingly, the same model formally applies to fermionic dark matter, although with a different interpretation. In the case of fermionic dark matter particles, one has to account for the pressure arising from the Pauli exclusion principle. For completely degenerate fermions, this pressure term can be computed from the Fermi-Dirac distribution at zero temperature, yielding (for spin-$1/2$ fermions in three dimensions) the equation of state  \cite{Chavanis2004} 
\begin{equation}
p= K \rho^{5/3} \quad \text{with} \quad
K= \left(\frac{3}{8 \pi} \right)^{2 / 3} \frac{(2 \pi \hbar)^{2}}{5 m^{8 / 3}}.  
\end{equation}
That is, a polytropic equation of state with a polytropic index $n=3/2$. This is indeed equivalent to an effective short-range potential $\Phi_{coll}=(5 / 2) K \rho^{2 / 3}$ in the Schrödinger equation\footnote{To the best of our knowledge, this has been first noticed in Ref. \cite{Manfredi} for quantum plasmas.} \cite{Chavanis2011},
\begin{equation}\label{ff}
i \hbar \frac{\partial \psi}{\partial t}=-\frac{\hbar^{2}}{2 m} \Delta \psi+m \Phi \psi+\frac{\left(3 \pi^{2}\right)^{2 / 3}}{2} N^{2 / 3} \frac{\hbar^{2}}{m}|\psi|^{4 / 3} \psi.
\end{equation}
Hence, apart from the difference in the exponent, the pressure term arising from the short-range self-interaction in bosonic dark matter is formally identical to the pressure term originating from the Pauli exclusion principle for fermionic dark matter particles. Taking advantage of this formal analogy, our model will consist in the quantum hydrodynamic set of equations for dark matter particles, with a polytropic pressure, coupled to the classical hydrodynamic equations for baryonic matter, and to the Poisson equation describing the gravitational potential in the weak-field regime. That is,
\begin{equation}\label{H}
\begin{aligned}
\frac{\partial \rho_b}{\partial t}&+\nabla \cdot(\rho_b \mathbf{u}_b) =0,\\
\frac{\partial \mathbf{u}_b}{\partial t}&+(\mathbf{u}_b \cdot \nabla) \mathbf{u}_b=-\nabla \Phi-\frac{K_b}{\rho_b}\nabla \rho_b^{\gamma_b}, \\
\frac{\partial \rho_d}{\partial t}&+\nabla \cdot(\rho_d \mathbf{u}_d)=0, \\
\frac{\partial \mathbf{u}_d}{\partial t}&+(\mathbf{u}_d \cdot \nabla) \mathbf{u}_d=-\nabla \Phi-\frac{K_d}{\rho_d} \nabla \rho_d^{\gamma_d}-\frac{1}{m_d} \nabla Q_d,\\
\Delta \Phi&=4 \pi G (\rho_b + \rho_d),
\end{aligned}
\end{equation}
where the subscripts $b$ and $d$ stand for baryonic and dark matter, respectively. This shall be our ultimate model. It covers both bosonic and fermionic dark matter, at the fluid level of description (see for instance \cite{Chavanis2019}). Besides, in the limit $\hbar \to 0$ (for a vanishing quantum potential $Q_d$), the model corresponds to classical dark matter (warm dark matter for $K_d \neq 0$ or pressureless CDM for $K_d=0$). The limit $\hbar \to 0$ also corresponds to BECDM in the TF approximation \cite{Chavanis2020} (where the
quantum potential can be neglected).  


\section{Jeans instability analysis}
We address in this section the mechanism of Jeans instability in the model (\ref{H}). We restrict ourselves to linear perturbations around a uniform non-expanding background. {For the zeroth-order dynamics, we consider a stationary, infinite,
homogeneous, and isotropic equilibrium fluid, and add small perturbations around these equilibrium values} 

\begin{equation}
\rho_b=\rho_{b0}+\delta \rho_b, \quad \rho_d=\rho_{d0}+\delta \rho_d, \quad \mathbf{u}_b=\mathbf{u}_{b0}+\delta \mathbf{u}_b, \quad \mathbf{u}_d=\mathbf{u}_{d0}+\delta \mathbf{u}_d, \quad \Phi=\Phi_{0}+\delta \Phi,
\end{equation}
where we can set $\mathbf{u}_{i0}=0$ ($i=b,d$) and $\Phi_0=0$. Inserting these quantities into Eq. (\ref{H}) and keeping only first-order terms, i.e., linearization, we get 
\begin{equation}\label{Hlin}
\begin{aligned}
\frac{\partial \delta \rho_b}{\partial t}&+ \rho_{b0}\nabla \cdot \delta \mathbf{u}_b=0, \\
\frac{\partial \delta \mathbf{u}_b}{\partial t}&=- {c_{b}^{2} } \frac{\nabla \delta \rho_b}{\rho_{b0}} -\nabla \delta\Phi, \\
\frac{\partial \delta \rho_d}{\partial t}&+\rho_{d0}\nabla \cdot \delta \mathbf{u}_d=0, \\
\frac{\partial \delta \mathbf{u}_d}{\partial t}&=-{c_{d}^{2} } \frac{\nabla \delta \rho_d}{\rho_{d0}}-\nabla \delta \Phi+\frac{\hbar^{2}}{4 m_d^{2} \rho_{d0}} \nabla(\Delta \delta \rho_d), \\
\Delta \delta \Phi&=4 \pi G  (\delta \rho_b+\delta \rho_d),
\end{aligned}
\end{equation}
where we have defined the sound velocities for the two media\footnote{By speaking of ‘sound velocity', one is tacitly assuming that the fluids involved are collisional, i.e., that there are considerable interactions between the particles comprising each matter component. Strictly speaking, a fluid description is not correct for collisionless media and the kinetic treatment is more appropriate in this case (see Appendix). Fluid-like equations can nonetheless be derived by taking velocity moments of the Vlasov or the Wigner equation, and identifying the velocity dispersion with $c_i$ ($i=b,d$). Hence, ‘sound velocities' should be understood as a generic expression for a velocity dispersion parameter.},
\begin{equation}
c_b^2 \equiv K_b \gamma_b \rho_{b0}^{\gamma_b-1} \quad \text{and} \quad c_d^2 \equiv K_d \gamma_d \rho_{d0}^{\gamma_d-1}.
\end{equation}
In particular, for bosonic dark matter, one has $c_d^2=4 \pi a \hbar^2 \rho_{d0}/m_d^3$, whereas for fermionic dark matter, one has $c_d^2=(3/ \pi)^{2/3} h^2 \rho_{d0}^{2/3}/12m_d^{8/3}$.
Above, we have used the so-called “Jeans
swindle”, i.e., by considering that the gravitational potential is sourced only by the density perturbations and not
by the density background $\rho_{i0}$ ($i=b,d$) (see for instance \cite{swindle} for
an elaborate discussion and formal justification). Expressing the perturbed quantities as plane waves $\propto \exp [i(\mathbf{k} \cdot \mathbf{r}-\omega t)]$, and combining the equations in Eq. (\ref{Hlin}), we obtain after simple manipulations the following dispersion relation

\begin{equation}\label{DR22}
1+ 4 \pi G \left[ \frac{\rho_{b0}}{\omega^2-c_b^2 k^2} + \frac{\rho_{d0}}{\omega^2-c_d^2 k^2- \hbar^2 k^4 /4 m_d^2} \right]=0.
\end{equation}
At this stage, one may check that, in the absence of dark matter (i.e., $\rho_{d0}=0$), the usual dispersion relation is recovered, namely,
\begin{equation}\label{drr}
\omega^2=-4 \pi G  \rho_{b0}+c_b^2 k^2.
\end{equation}
Above, two cases have to be distinguished: For $\omega^2>0$, the angular frequency $\omega$ is real and the perturbation behaves
with time as $\operatorname{e}^{-i \omega t}$, i.e., an oscillatory regime with a frequency $\pm \sqrt{\omega^2}$, while for $\omega^2<0$, the frequency
is imaginary ($\omega = i \gamma$) and the perturbation evolves exponentially
with time with a rate $\gamma$ (with $\gamma= \pm \sqrt{- \omega^2}$). In this case, there is a growing mode and a decaying
mode; The growing mode is responsible for the Jeans instability. The critical wave number $k_J$ separating between the oscillatory regime and the unstable regime is known as the Jeans wave number.
It can be inferred from the dispersion relation (\ref{drr}) by setting $\omega^2=0$. It follows as
\begin{equation}
k_J= \sqrt{\frac{4 \pi G \rho_{b0}}{c_b^2}}.
\end{equation}
The presence of a dark matter background modifies the dispersion relation, modifying therefore this critical value. By setting $\omega^2=0$ in the dispersion relation (\ref{DR22}), one obtains the critical wave number $k^{*}$, separating between stable and unstable modes. {Setting $\omega^2=0$ in Eq. (\ref{DR22}) and solving for $k^2$, one finds two solutions (one positive and one negative)}

\begin{equation}
k^2_{\pm}=\frac{-c_b^2 c_d^2 + \pi G \hbar^2 \rho_{b0}/m_d^2   \pm  \sqrt{(c_b^2 c_d^2+ \pi G \rho_{b0} \hbar^2 / m_d^2)^2  + 4 \pi G \rho_{d0} c_b^4 / m_d^2}}{c_b^2 \hbar^2/2m_d^2}.
\end{equation}
{Keeping the positive solution, we obtain the critical wave number $k^{*}$, separating between stable and unstable modes. In a dimensionless form, it reads}

\begin{equation}\label{Jeans}
 \frac{k^{*2}}{ k_J^2} =\frac{1}{2} +\frac{m_d^2}{2 \pi G \hbar^2 \rho_{b0}} \left [-c_b^2 c_d^2 + \sqrt{(c_b^2 c_d^2+ \pi G \hbar^2 \rho_{b0}/m_d^2)^2+4 \pi G \hbar^2 c_b^4\rho_{d0}/m_d^2} \right].
\end{equation}
One may easily check that in the absence of a dark matter background ($\rho_{d0}=0$), one has $k^{*}=k_J$ as expected. 
Two limiting cases of Eq. (\ref{Jeans}) are worth a closer examination.

\subsection{The limit $\hbar \to 0 $}
At one extreme, one may study the limit $\hbar \to 0$ (or better a vanishing quantum potential $Q_d$). {red}{It should be noted that the formal limit $\hbar \to 0$ is taken here assuming that $c_d^2$ remains constant. This limiting case corresponds to classical dark matter particles, where quantum effects are ignored in the first place and $c_d^2$ has a classical (thermal) origin. It may also apply to BECDM in the so-called Thomas-Fermi (TF) approximation, where the quantum potential can be ignored (see discussion in the next section).} In this limit, the critical wave number (\ref{Jeans}) reduces to

\begin{equation}\label{kJc}
\frac{k^{*2}}{k_J^2}= 1+ \frac{c_b^2}{c_d^2}\frac{\rho_{d0}}{\rho_{b0}}.
\end{equation}
This limit corresponds to the critical wave number derived by Kremer \textit{et al.} \cite{Jkin3,Kremer1} for classical dark matter through a kinetic approach, using collisionless Boltzmann (Vlasov) equations, upon identifying the sound velocities $c_i$ ($i=b,d$) with velocity dispersions. Eq. (\ref{kJc}) shows that the presence of dark matter tends to increase the critical wave number, rendering the system unstable for smaller wavelengths. Note however that, if Eq. (\ref{kJc}) applies to BECDM in the TF approximation, the effect may go in the opposite direction if the self-attraction between the bosons is attractive, since in this case one has $c_d^2<0$ (see for instance \cite{Chavanis2020}).

It may be instructive to note that, taking further the limit $c_d \to 0$, one has $k^{*}\to \infty$, which corresponds to a vanishing critical wavelength $\lambda^{*}:=2\pi / k^{*}$. That is, the whole system is unstable because dark matter does not oppose pressure of any kind to gravity.

\subsection{The limit {$c_d^2=0$} }
At the other extreme, one retains only the effect of quantum pressure for dark matter. This limit corresponds to BECDM without short-range self-interaction (see for instance \cite{Chavanis2020,Ourabah2020bis}). In this case, the stability of dark matter is assured solely by the quantum pressure acting against gravity, and the critical wave number (\ref{Jeans}) becomes
\begin{equation}\label{eq25}
\frac{k^{*2}}{ k_J^2} =\frac{1}{2} + \sqrt{\frac{1}{4}+\frac{c_b^4 m_d^2 \rho_{d0}}{\pi G \hbar^2 \rho_{b0}^2}}.
\end{equation}
The latter combines thermal effects for baryonic matter and quantum effects for dark matter particles. It is clear from the above relation that the presence of a dark matter background increases the critical wave number $k^{*}$. This means that a dark matter background increases the instability region, making the medium unstable for smaller wavelengths.


{In what follows, we shall compare the (in)stability conditions (\ref{kJc}) and (\ref{eq25}) with astrophysical data of Bok globules. For the ratio of densities $\rho_{d0}/\rho_{b0}$, we take the ratio of the density parameter $\Omega_d/ \Omega_b$ today. In fact, in the Jeans analysis, one assumes from the start an \textit{infinite} and \textit{homogeneous} medium with a constant density for the zeroth order dynamics \cite{Binney}. Hence, the densities $\rho_{i0}$ ($i=b,d$) are understood as global densities (see e.g. \cite{Kremer1}). We are then left with a single parameter to be fixed upon confronting the stability criteria with the data, namely $c_b/c_d$ for Eq. (\ref{kJc}) and the mass $m_d$ for Eq. (\ref{eq25}).}

\section{Analysis of Bok globules}
In this section, {we compare the stability criteria (\ref{kJc}) and (\ref{eq25})} with astrophysical stability observations. For that purpose, instead of the Jeans wave number, it is more appropriate to define a critical mass, i.e., the Jeans mass, which corresponds to the mass initially contained in a sphere of diameter $\lambda^*:=2 \pi / k^*$, that is 
\begin{equation}
M^{*} \equiv \frac{4 \pi \rho_{0}}{3}\left(\frac{\lambda^{*}}{2}\right)^{3},
\end{equation}
where, in our scenario, $\rho_0$ accounts for both baryonic and dark matter contents, i.e., $\rho_0 \equiv \rho_{b0}+\rho_{d0}$.

We are particularly interested here in confronting the obtained critical mass, in the presence of dark matter, with the observed stability of Bok globules. The latter are nearly isolated and simple-shaped clouds of interstellar gas and dust, with core temperatures of the order of $10K$ and masses around $10M_{\odot}$, that can experience star formation \cite{Kandori}. Besides, Bok globules have masses of the same order of magnitude as their corresponding Jeans mass; hence a small correction to the Jeans mass may lead to a different prediction for their stability. This last feature places Bog globules as excellent laboratories to test different predictions on the Jeans mechanism. 

In Table \ref{tab:globules}, we reproduce the kinetic temperature, particle number density, mass, Jeans mass, and observed stability for several Bok globules given in Ref. \cite{Kandori}. One may observe that 7 of the Bok globules therein have a mass smaller than their corresponding Jeans mass (hence, they are predicted as stable), yet the observation reveals that they do experience star formation. This discrepancy between prediction and observation has been studied from various angles in the literature, and several solutions have been suggested based on modified gravity theories \cite{Vainio,Claudio}, nonequilibrium thermodynamics \cite{Oursup}, and generalized uncertainty principles \cite{gup1,gup2}. We explore here the possibility that the presence of a dark matter background can provide an explanation to this discrepancy. We note in passing that, as pointed out recently \cite{Gho}, the presence of dark matter may explain recent observations showing that some Bok globules are prolate in some regions instead of spherical.

To proceed, one may observe from table \ref{tab:globules} that a correct stability for all Bok globules is achieved if the Jeans mass is reduced by a factor $2/5$ (as first noticed in \cite{Claudio}). This is illustrated in Fig. \ref{fpleiadesy}, reporting the mass and the Jeans mass for 7 of the Bok globules of Table \ref{tab:globules} whose predicted stability is contradicted by observation. Considering the critical wave numbers (\ref{kJc}) and (\ref{eq25}), we may explore to what extent the presence of a dark matter background may explain these discrepancies. {We first consider the criterion (\ref{kJc}), corresponding to classical dark matter particles or BECDM in the TF approximation.} In this case, the critical wave number and the corresponding critical mass depend only on the ratios $\rho_{d0}/\rho_{b0}$ and $c_d/c_b$. For the first ratio, following Ref. \cite{Kremer1}, we take the ratio of the density parameter $\Omega_d/ \Omega_b$ today, i.e., $\rho_{d0}/\rho_{b0} \approx 5.5$ \cite{Olive}, as it has not changed that much during the evolution of the Universe. We are then left with a single dimensionless parameter, namely $c_d/c_b$. Using the critical wave number (\ref{kJc}), the critical mass reads as
\begin{equation}
\frac{M^*}{M_J}= \frac{\rho_{b0}+\rho_{d0}}{\rho_{b0}} \left (1+\frac{\rho_{d0}}{\rho_{b0}}\frac{c_b^2}{c_d^2} \right)^{-3/2},
\end{equation}
where $M_J$ is the usual Jeans mass, in the absence of dark matter. This critical mass $M^*$ accounts for both baryonic (visible) and dark matter. At this stage, one has to note that the reported Bok globule masses are obtained through spectroscopic methods and account only for visible matter. One has to account for that and compare the Bok globule mass $M$ to $\bar{M}$ such that $\bar{M}=M^{*}\rho_{b0}/ \rho_0$, where $\rho_0 = \rho_{b0}+\rho_{d0}$ is the total mass density. That is,
\begin{equation}\label{MMJ}
\frac{\bar{M}}{M_J}=\left(1+\frac{\rho_{d0}}{\rho_{b0}}\left(\frac{
c_b^2}{c_d^2}\right)\right)^{-3/2}\approx \left(1+5.5 \left(\frac{c_b^2}{c_d^2} \right)\right)^{-3/2}.
\end{equation}
This is equivalent to comparing the total mass of the Bok globule (including the dark matter content) to the critical mass $M^*$. Using Eq. (\ref{MMJ}), we give in Table \ref{tab:globulesFR} the lower bounds for $c_b/c_d$ in order to match the observed instability for each Bok globule. This is also illustrated in Fig. \ref{fpleiadesy} showing the critical mass (\ref{MMJ}) as a function of $c_b/c_d$, where we have highlighted the two extreme Bok globules, namely CB131 and CB184 along with the sufficient condition $\bar{M}/M_J=2/5$. One may appreciate that the lower bounds on $c_b/c_d$ are of the same order of magnitude of the ratio of velocity dispersions known in the literature. In fact, numerical simulations on Milky Way-like galaxies,
including baryonic and dark matter, provide \cite{Olive} $c_b/c_d=93/170 \approx 0.547$, while using Eq. (\ref{MMJ}), the critical value $\bar{M}/M_J=2/5$ is achieved for $c_b/c_d \approx 0.391$, which represents a relative deviation of $\sim 28 \%$.

{At the other extreme, we consider the case (\ref{eq25}) where only quantum effects are retained. The critical mass $\bar{M}$ reads then as}

\begin{equation}
\frac{\bar{M}}{M_J}=\left(\frac{k^*}{k_J}\right)^{-3},
\end{equation}
{where $k^*$ is given by Eq. (\ref{eq25}). This limit corresponds to BECDM without short-range interaction (i.e., $c_d^2=0$). The BECDM model is particularly relevant for our analysis since Bok globules are among the coldest known astrophysical objects. Computing the critical mass $\bar{M}$ for the Bok globules considered here, we obtain for each Bok globule the lower mass $m_d$ that allows accounting for the data. This is shown in Table \ref{tab:globulesFR}. One may observe that, when only quantum effects are retained, the model adequately accounts for the data for ultralight dark matter particles, with masses of the order $m_d \sim 10^{-18}-10^{-17}eV/c^2$. This can be compared with the usual predictions of the BECDM model. For noninteracting bosons, in order to reproduce the
scales of dark matter halos, the dark matter particle is estimated as \cite{Chavanis2019} $m_d \sim 10^{-22}eV/c^2$, which is three to four order of magnitude smaller than our predictions. However, when a self-interaction
between the bosons is allowed, a large mass window is
open, allowing for much more massive bosons \cite{Chavanis2019}; the mass of the bosons may range from $10^{-22}eV/c^2$
to a few $eV/c^2$}
{(this may be particularly relevant since masses around $m_d \sim 10^{-22}-10^{-21}eV/c^2$ are in tension
with observations of the Lyman-$\alpha$ forest \cite{Hui}). We note in passing that bosons with a mass around $10^{-17}eV/c^2$ have been recently proposed \cite{Torres,Guzman} as constituents of stable boson stars that could mimic supermassive black holes.}

{To study the effect introduced by self-interaction in the BECDM model and how this allows for much larger dark matter particle masses, let us discuss briefly the validity of each limit ($\hbar \to 0$, i.e. Eq. (\ref{kJc}) and $c_d^2=0$, i.e. Eq. (\ref{eq25})) for BECDM. From the condition of hydrostatic equilibrium, one may define the following dimensionless parameter \cite{Chavanisbook}}
\begin{equation}
\chi \equiv \frac{G M^{2} m_d|a|}{\hbar^{2}},
\end{equation}
{$M$ being the total mass of the self-gravitating BECDM halo. For $\chi \ll 1$, we are in the non-interacting limit in which scattering is negligible. In that case, the equilibrium results from the interplay
between gravitational attraction and quantum pressure (cf. Eq. (\ref{eq25})). On the contrary, for $\chi \gg 1$, we are in the TF limit in which the quantum potential is negligible. In that case, the equilibrium is ensured by the balance
between gravitational attraction and repulsive scattering (for $g > 0$). The Jeans mass in this case corresponds formally to that of classical dark matter, with sound velocity given by 
$c_d^2=4 \pi a \hbar^2 \rho_{d0}/m_d^3$. Using the values of $c_b/c_d$ given in Table \ref{tab:globulesFR} to account for the data and the sound velocity for BECDM, one may deduce a relationship between the mass $m_d$ and the scattering length $a$. For the Bok globules considered in Table \ref{tab:globulesFR}, we find}

\begin{equation}\label{mppp} 
\frac{m_d}{1 \mathrm{eV} / c^{2}} \sim 10 \left(\frac{a}{1 \mathrm{fm}}\right)^{1 / 3},
\end{equation}
{confirming that, in the presence of short-range interactions, much larger boson masses are allowed. In fact, considering the typical value $a =10^{-6}fm$ of the scattering length
observed in laboratory BEC experiments \cite{exp}, Eq. (\ref{mppp}) yields $m_d \sim 10 eV/c^2$, which is, as expected, much larger than the boson mass predicted in the noninterracting case.}
 
\begin{table}[h!]
\scriptsize
\begin{tabular}{|l|c|c|c|c|c|c|c|} \hline
Bok Globule &$T\text{[K]}$ & $n\text{[cm$^{-3}$]}$ & $M [M_{\odot}]$  & $M_J [M_{\odot}]$ &  Stability \\ \hline
CB 87	& 11.4 &	$(1.7\pm 0.2)\times 10^4$	& $2.73\pm 0.24$ & 9.6 &  stable \\ \hline
CB 110 & 21.8 & $(1.5\pm 0.6)\times 10^5$ & $7.21\pm 1.64$ & 8.5  &  unstable \\ \hline
CB 131 & 25.1 & $(2.5\pm 1.3)\times 10^5$ & $7.83\pm 2.35$ & 8.1  &  unstable \\ \hline
CB 134 & 13.2 & $(7.5\pm 3.3)\times 10^5$ & $1.91\pm 0.52$ & 1.8  &  unstable \\ \hline
CB 161 & 12.5 & $(7.0\pm 1.6)\times 10^4$ & $2.79\pm 0.72$ & 5.4   &  unstable \\ \hline
CB 184 & 15.5 & $(3.0\pm 0.4)\times 10^4$ & $4.70\pm 1.76$ & 11.4  &  unstable \\ \hline
CB 188 & 19.0 & $(1.2\pm 0.2)\times 10^5$ & $7.19\pm 2.28$ & 7.7 &  unstable \\ \hline
FeSt 1-457 & 10.9 & $(6.5\pm 1.7)\times 10^5$ & $1.12\pm 0.23$ & 1.4  &  unstable \\ \hline
Lynds 495 & 12.6 & $(4.8\pm 1.4)\times 10^4$ & $2.95\pm 0.77$ & 6.6  &  unstable \\ \hline
Lynds 498 & 11.0 & $(4.3\pm 0.5)\times 10^4$ & $1.42\pm 0.16$ & 5.7 & stable \\ \hline
Coalsack & 15 & $(5.4\pm 1.4)\times 10^4$ & $4.50$ & 8.1 & stable \\ \hline
\end{tabular}
\caption{Kinetic temperature, particle number density, mass, Jeans mass, and observed stability for several Bok globules \cite{Kandori}.}
\label{tab:globules}
\end{table}

\begin{table}[h!]
\scriptsize
\begin{tabular}{|l|c|c|c|c|c|c|c|} \hline
Bok Globule &$T\text{[K]}$ & $n \text{[cm$^{-3}$]}$ & $M [M_{\odot}]$  & $M_J [M_{\odot}]$ & $(c_b/c_d)_{min}$ & {$m_d$ [$10^{-17}eV/c^2$]} \\ \hline
CB 110 & 21.8 & $(1.5\pm 0.6)\times 10^5$ & $7.21\pm 1.64$ & 8.5  & 0.145 & {$1.248 $ } \\ \hline
CB 131 & 25.1 & $(2.5\pm 1.3)\times 10^5$ & $7.83\pm 2.35$ & 8.1  & 0.064 & {$0.343 $ } \\ \hline
CB 161 & 12.5 & $(7.0\pm 1.6)\times 10^4$ & $2.79\pm 0.72$ & 5.4   & 0.317 & {$3.838 $ }\\ \hline
CB 184 & 15.5 & $(3.0\pm 0.4)\times 10^4$ & $4.70\pm 1.76$ & 11.4  & 0.383 & {$2.636 $ } \\ \hline
CB 188 & 19.0 & $(1.2\pm 0.2)\times 10^5$ & $7.19\pm 2.28$ & 7.7 & 0.092 & {$0.789 $ } \\ \hline
FeSt 1-457 & 10.9 & $(6.5\pm 1.7)\times 10^5$ & $1.12\pm 0.23$ & 1.4  & 0.171 & {$6.252 $ } \\ \hline
Lynds 495 & 12.6 & $(4.8\pm 1.4)\times 10^4$ & $2.95\pm 0.77$ & 6.6  & 0.359 & {$3.751 $} \\ \hline
\end{tabular}
\caption{Kinetic temperature, particle number density, mass, and Jeans mass for 7 of the Bok globules of Table \ref{tab:globules}, whose predicted stability is contradicted by observation, together with the saturation bounds for $c_b/c_d$ obtained with Eq. (\ref{kJc}) {and for $m_d$ obtained with Eq. (\ref{eq25})}.}
\label{tab:globulesFR}
\end{table}

\begin{center}
\begin{figure}
 \includegraphics[height=80mm]{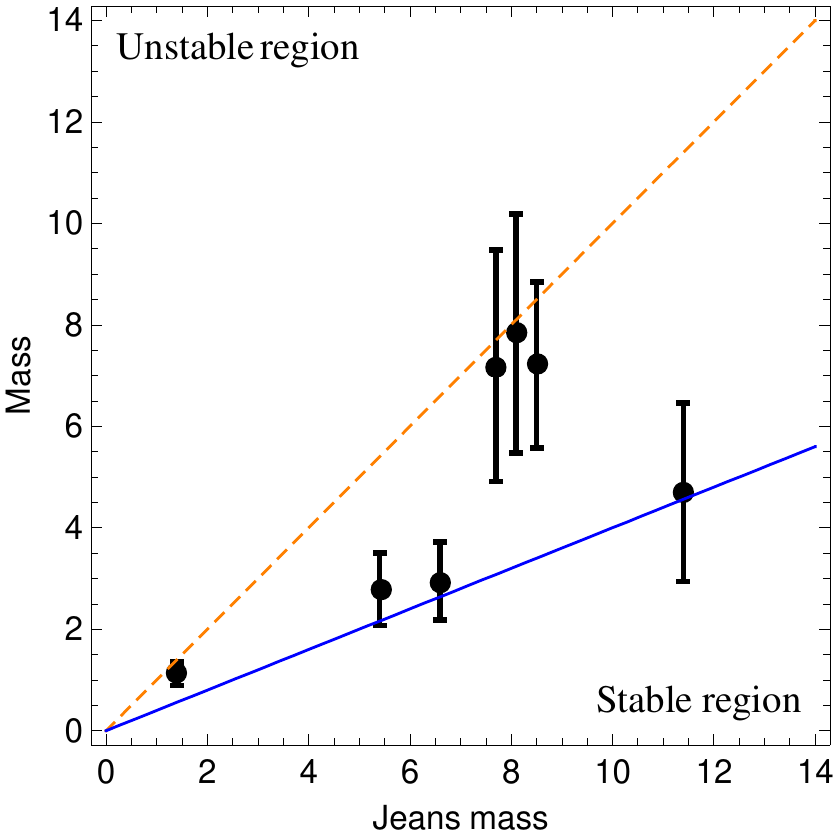}
\caption{Mass and Jeans mass of 7 of the Bok globules given in Table \ref{tab:globules} whose predicted stability is contradicted by observation. The dashed line delimits between the stable region and the collapsing region for the usual Jeans criterion (in the absence of dark matter). The solid line represents the bound $2/5$ given in Ref. \cite{Claudio}, as a sufficient condition to account for the data.  }\label{fpleiadesy}
\end{figure}
\end{center}

\begin{center}
\begin{figure}
 \includegraphics[height=80mm]{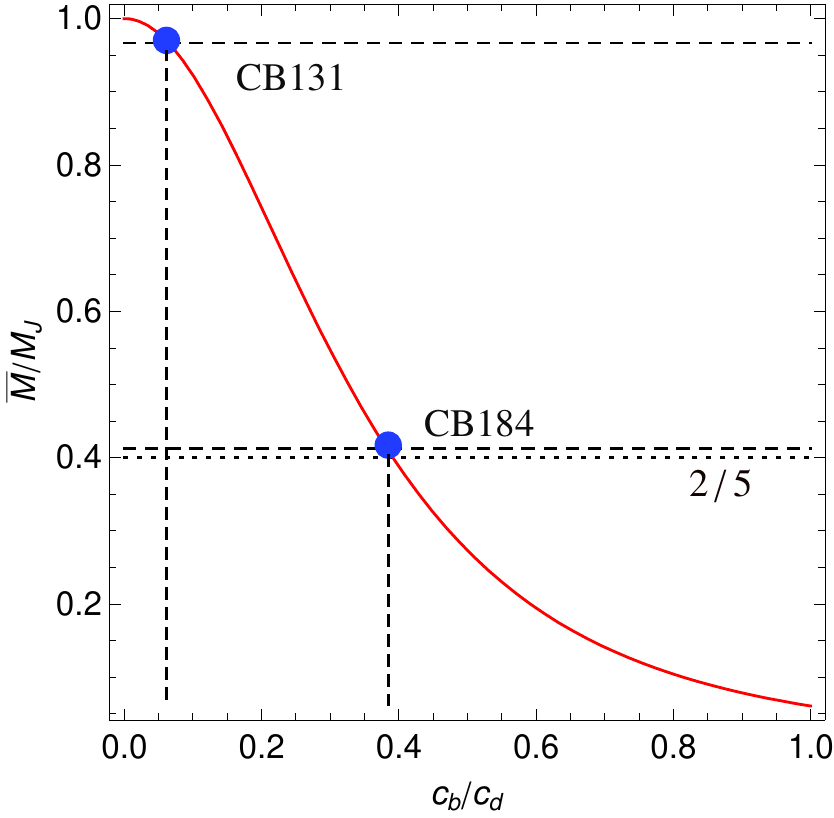}
\caption{$\bar{M}/M_J$ as a function of $c_b/c_d$. We highlight the two extreme cases of Bok globules given in Table \ref{tab:globulesFR}, namely CB131 and CB184. The horizontal dotted line represents the bound $2/5$ given in Ref. \cite{Claudio}.  }\label{fpleiadesy}
\end{figure}
\end{center}

\section{Conclusion}
In this paper, we have presented a fluid-like model for a self-gravitating medium composed of two species, namely baryonic (visible) and dark matter particles. Baryonic matter is treated classically (which is appropriate for most astrophysical situations) while dark matter is treated through a quantum hydrodynamic model with a polytropic equation of state. The main virtue of this model is that it covers both fermionic dark matter, e.g., massive neutrinos, and bosonic dark matter, e.g., axions, while it reproduces classical warm or cold dark matter in the proper limits. 

Through the model, we have studied the mechanism of Jeans instability, establishing general stability criteria and discussing their relevant limits. In the simplest case of classical dark matter, we have shown that the effect of a dark matter background on the critical Jeans mass depends solely on the ratios of mass densities and velocity dispersions. Exploiting that, we have confronted the model with observed stability data of Bok globules and have shown that the model adequately accounts for the data for dark matter parameters of the same order of magnitude as those predicted independently from numerical simulations. {At the other extreme, where only quantum effects are retained, we have shown that the model accounts for the data for ultralight dark matter particles with masses $\sim 10^{-18}-10^{-17}eV/c^2$.}

This work may open up new prospects for research in the near future. In particular, a closer examination of Bok globule stability observations in more general cases seems worthwhile. 

\appendix
\section*{Appendix: Kinetic treatment}

In the main text, we restrict ourselves, for simplicity, to the fluid level of description. The problem may nonetheless be equally well formulated in a kinetic way. We provide below the basic steps for such a description.

In a kinetic approach, one considers the phase space spanned by the space and velocity (or momentum) coordinates. A state of the given (classical) system is characterized by the one-particle distribution function $f(\mathbf{r},\mathbf{v};t)$, whose space-time evolution in the phase space is given by the Boltzmann equation.

In our scenario, we treat baryonic matter classically and ignore the effect of collisions. The evolution of the one-particle distribution function then follows the collisionless Boltzmann (Vlasov) equation. That is,
\renewcommand{\thesection}{A.\arabic{section}}
\setcounter{equation}{0}
\begin{equation}\label{Boltzmann}
\frac{\partial f_b}{\partial t}+\mathbf{v}_b \cdot \nabla f_b-\nabla \Phi \cdot \frac{\partial f_b}{\partial \mathbf{v}_b}=0.
\end{equation}
For dark matter, one needs a similar kinetic equation that stays valid in the quantum regime. This can be done upon defining the Wigner function\footnote{Formally, the Wigner
function is not a \textit{bona fide} distribution and should
be rather regarded as a quasi-distribution, since it can
take negative values. It is nevertheless a very
useful mathematical tool to study the collective behavior of quantum systems \cite{ga1,Ourrr}.}
\begin{equation}\label{Wf}
f_d(\mathbf{r}, \mathbf{p}_d; t)= \frac{1}{(2 \pi \hbar)^3} \int d \mathbf{y} \exp (i \mathbf{p}_d \cdot \mathbf{y} / \hbar^3)\psi^{*}(\mathbf{r}+\mathbf{y} / 2, t) \times \psi(\mathbf{r}-\mathbf{y} / 2, t),
\end{equation}
where $\mathbf{p}_d \equiv m_d \mathbf{v}_d$ is the dark matter particle momentum. The Wigner function (\ref{Wf}) is simply the Fourier transform of the auto-correlation function corresponding to the wave-function $\psi$. It is normalized here such that
\begin{equation}
\int f_d(\mathbf{r}, \mathbf{p}_d; t)  d \mathbf{p}=\left|\psi(\mathbf{r}, t)\right|^{2} = \rho_d(\mathbf{r},t),
\end{equation}
where $\rho_d(\mathbf{r},t)$ denotes the dark matter mass density. Following the so-called Wigner-Moyal procedure \cite{Wigner,Moyal}, one may write the Schrödinger equation (\ref{GP}) in the form of a kinetic equation as follows (see for instance \cite{Tito} for detailed calculations)

\begin{equation}\label{WW}
i\hbar\left(\frac{\partial}{\partial t}+\frac{\textbf{p}_d}{m_d}\cdot\nabla\right)f_d\left(\textbf{r},\textbf{p}_d;t\right)=\int \Phi_{\textbf{k}} \Delta f_d \exp{\left(i\textbf{k}\cdot\textbf{r}\right)}d\textbf{k},
\end{equation}
where $\Delta f_d:=\left[f_d^{-}-f_d^{+}\right]$, with $f_d^{\pm}:=f_d(\mathbf{r}, \mathbf{p}_d \pm \hbar \mathbf{k} / 2m_d; t)$, and
\begin{equation}
\Phi_{\textbf{k}}:= m_d \iint (\Phi + \Phi_{coll}) f_d\left(\textbf{r},\textbf{p}_d;t\right)d\textbf{p}_d\exp{\left(-i\textbf{k}\cdot\textbf{r}\right)}d\textbf{r}.
\end{equation}
It is interesting to observe that, although describing a quantum system, Eq. (\ref{WW}) has the mathematical structure of a classical master equation. Besides, in the limit $\hbar \to 0$, it reduces to the classical Vlasov equation (\ref{Boltzmann}).

In addition to the kinetic equations (\ref{Boltzmann}) and (\ref{WW}), the Poisson equation reads as
\begin{equation}\label{PP}
\Delta \Phi = 4 \pi G (\rho_b + \rho_d) =4 \pi G \left(\int f_b d \mathbf{v}_b + \int f_b d \mathbf{p}_d \right).
\end{equation}
Eqs. (\ref{Boltzmann}), (\ref{WW}), and (\ref{PP}) constitute the kinetic counterpart of the model (\ref{H}) discussed in the main text. Following the standard steps, i.e., considering small perturbations around the equilibrium values represented by plane waves, and making use of the Jeans swindle (see e.g., \cite{Ourr}), one arrives at the following dispersion relation

\begin{equation}\label{DRK}
 1+ \frac{4 \pi G}{k^2} \int_{- \infty}^{\infty} \frac{\partial f_{b0} / \partial v_b }{v_b - \omega / k} d v_b + \left(\frac{4 \pi G}{k^2}-g \right)\frac{m_d}{\hbar} \int_{- \infty} ^{\infty} \frac{f_{d0}(p_d + \hbar k/2)- f_{d0}(p_d - \hbar k/2) dp_d}{p_d k/m_d- \omega}=0,
\end{equation}
where we have considered, without loss of generality the wave vector $\mathbf{k}$ to be parallel to the $x$-axis and have redefined the equilibrium distributions as the projected (marginal) distributions along that axis. That is,

\begin{equation}
f_{b0} \to \iint f_{b0} d \mathbf{v}_{b \perp} \quad \text{and} \quad f_{d0} \to \iint f_{d0} d \mathbf{p}_{d \perp},
\end{equation}
with $v_d$ and $p_d$ representing the components of the velocity and momentum along the $x$-axis. Equation (\ref{DRK}) is the kinetic counterpart of the dispersion relation (\ref{DR22}), to which it reduces in the proper limit. To show that, we consider an even distribution, e.g., a Maxwellian distribution, for baryonic matter, which is characteristic of equilibrium or nearly equilibrium situations, while for dark matter we identify $f_{d0}$ with a Dirac delta, i.e., $f_{d0}= \rho_{d0} \delta (p)$, since we are considering dark matter at $T=0$. With these assumptions, Eq. (\ref{DRK}) reduces to Eq. (\ref{DR22}), obtained in the hydrodynamic formulation, with $c_b^2 \to \langle v_b^2 \rangle$ and $c_d^2=g \rho_0$. The same lines of reasoning can be applied to Eq. (\ref{ff}), for fermionic dark matter particles, leading to Eq. (\ref{ff}), with $c_d^2=(3/ \pi)^{2/3} h^2 \rho_{d0}^{2/3}/12m_d^{8/3}$. It may be interesting to note that, in the limit $\hbar \to 0$ and $g=0$, Eq. (\ref{DRK}) reduces to the dispersion relation derived by Kremer \textit{et al.} \cite{Jkin3,Kremer1} for classical dark matter in the kinetic picture. Note however that the kinetic treatment is not strictly equivalent to the fluid approach. The former is in some sense more general as it allows studying purely kinetic effects, such as the Landau damping, while the fluid approach does not.

\end{document}